\documentclass[fontsize=11pt]{article}
\usepackage[T1]{fontenc}
\usepackage[utf8]{inputenc}
\usepackage{geometry}
\usepackage{lineno,hyperref}
\usepackage{newtxtext}
\usepackage{amsmath,amssymb,mathtools}
\usepackage{graphicx}
\usepackage{caption}
\usepackage{authblk}

\hypersetup{
	colorlinks=true,
	linkcolor=blue,
	linkbordercolor={0 0 0},
	citecolor=blue,
}

\begin{document}
\title{Particle identification capability of a homogeneous calorimeter composed of oriented crystal}

\author[1,2]{Pietro Monti-Guarnieri\footnote{Corresponding author. Any correspondence should be sent to \url{pietro.monti-guarnieri@phd.units.it}}}
\author[3]{Laura Bandiera}
\author[3]{Nicola Canale}
\author[4,5]{Stefano Carsi}
\author[6,7]{Davide De Salvador}
\author[8,3]{Vincenzo Guidi}
\author[9]{Viktar Haurylavets}
\author[4,5]{Giulia Lezzani}
\author[1,2]{Francesco Longo}
\author[3]{Lorenzo Malagutti}
\author[4,5]{Sofia Mangiacavalli}
\author[8,3]{Andrea Mazzolari}
\author[10]{Matthew Moulson}
\author[8,3]{Riccardo Negrello}
\author[3]{Gianfranco Paternò}
\author[4,5]{Leonardo Perna}
\author[4,5]{Christian Petroselli}
\author[4,5]{Michela Prest}
\author[8,3]{Marco Romagnoni}
\author[4,5]{Giosuè Saibene}
\author[4,5]{Alessia Selmi}
\author[6,7]{Francesco Sgarbossa}
\author[10]{Mattia Soldani}
\author[3]{Alexei Sytov}
\author[9]{Victor Tikhomirov}
\author[5]{Erik Vallazza}

\affil[1]{Università degli Studi di Trieste, Trieste, Italy}
\affil[2]{INFN Sezione di Trieste, Trieste, Italy}
\affil[3]{INFN Sezione di Ferrara, Ferrara, Italy}
\affil[4]{Università degli Studi dell'Insubria, Como, Italy}
\affil[5]{INFN Sezione di Milano Bicocca, Milano, Italy}
\affil[6]{Università degli Studi di Padova, Padova, Italy}
\affil[7]{INFN Laboratori Nazionali di Legnaro, Legnaro, Italy}
\affil[8]{Università degli Studi di Ferrara, Ferrara, Italy}
\affil[9]{Institute of Nuclear Problems of the Belarusian State University, Minsk, Belarus}
\affil[10]{INFN Laboratori Nazionali di Frascati, Frascati, Italy}

\date{}
\maketitle

\begin{abstract}
	Recent studies have shown that the electromagnetic shower induced by a high-energy electron, positron or photon incident along the axis of an oriented crystal develops in a space more compact than the ordinary. On the other hand, the properties of the hadronic interactions are not affected by the lattice structure. This means that, inside an oriented crystal, the natural difference between the hadronic and the electromagnetic shower profile is strongly accentuated. Thus, a calorimeter composed of oriented crystals could be intrinsically capable of identifying more accurately the nature of the incident particles, with respect to a detector composed only of non-aligned crystals. Since no oriented calorimeter has ever been developed, this possibility remains largely unexplored and can be investigated only by means of numerical simulations. In this work, we report the first quantitative evaluation of the particle identification capability of such a calorimeter, focusing on the case of neutron-gamma discrimination. We demonstrate through Geant4 simulations that the use of oriented crystals significantly improves the performance of a Random Forest classifier trained on the detector data. This work is a proof that oriented calorimeters could be a viable option for all the environments where particle identification must be performed with a very high accuracy, such as future high-intensity particle physics experiments and satellite-based $\gamma-$ray telescopes. \\
	
	\noindent \emph{Keywords: Particle identification methods, Calorimeters, Calorimeter methods, Performance of High Energy Physics Detectors}
\end{abstract}

\section{Introduction}
In particle and astroparticle physics the measurement of the energy of an ultra-relativistic electron, positron or photon is usually performed using an electromagnetic calorimeter (eCAL). Traditionally, calorimeters are divided in two classes: they can be either homogeneous or sampling, depending on their design~\cite{gianotti2003}. In the last few years, a new hybrid layout has been developed: a homogeneous eCAL with longitudinal segmentation, also known as semi-homogeneous calorimeter~\cite{paesani2022,cantone2023}. With this name we identify a detector obtained by concatenating multiple layers of high-Z scintillating or Cherenkov crystals, each one coupled to a photodetector, without passive layers. The key feature of this design is the possibility to sample with a fine granularity the spatial development of the shower induced by the incident particles. This is a critical property for achieving an accurate background rejection level in a high-intensity radiation environment: for instance, heavy hadrons can be easily discriminated from lighter particles (such as photons and e$^\pm$), thanks to the difference in width and symmetry of the shower profile. Detectors based on this concept have been proposed both as barrel calorimeter for the Muon Collider instrumented beamline~\cite{ceravolo2022,ceravolo2022tris} and as Small Angle Calorimeter (SAC) for KLEVER, the third phase of the High Intensity Kaon Experiments (HIKE) project~\cite{moulson2019,moulson2023,gil2022,hike2023}. 

Several techniques can be used to increase the Particle Identification (PID) performance of a calorimeter, such as the measurement of the space-time development of the showers and the pulse-shape analysis~\cite{benaglia2016,badala2022}. A particularly novel and intriguing possibility is the use of oriented crystals: in fact, it has been known since the 1960s that the lattice structure of a crystal can modify the electromagnetic (e.m.) processes occurring inside it~\cite{kumakhov1977,uggerhoj2005}. In particular, if an ultra-relativistic e$^\pm$ impinges on the axis of an oriented crystal, the field generated by the lattice strings and seen in the particle frame of reference is boosted due to the Lorentz length contraction: in this way, the ``effective'' (perceived) field is enhanced and thus the bremsstrahlung (BS) cross-section increases~\cite{sorensen1987_nature,sorensen1996_notes}. If the energy of the incident particle is large enough, the effective field can reach an amplitude larger than the Schwinger critical field of QED ($\text{E}_0 \sim 1.32 \cdot 10^{16}$~V/cm), the threshold above which non-linear QED effects are observed in the vacuum. This is the so-called Strong Field (SF) regime: in this condition, e$^\pm$ emit hard synchrotron-like radiation, with a cross-section significantly larger than the standard bremsstrahlung value~\cite{uggerhoj2005,baryshevski1983,baier1998_book}. A similar effect is observed also for the electron-positron pair production (PP)~\cite{kimball1984a,baryshevski1989}: the combination of these effects results in a spatial acceleration of the the electromagnetic shower~\cite{soldani2024}. In order to observe the SF regime, two conditions must be satisfied~\cite{uggerhoj2005,baier1998_book,sorensen1996_notes}: 
\begin{enumerate}
	\item The angle between the direction of particle incidence and the crystal axis (``misalignment angle'', $\theta_{\text{mis}}$) must be smaller than a critical value~\cite{baier1998_book}:
	\begin{equation} \label{eq:1}
		\theta_{\text{mis}} < \Theta_0 = \frac{\text{U}_0}{\text{mc}^2}
	\end{equation}
	where $\text{mc}^2 = 511$~keV is the electron rest mass energy and U$_0$ is the potential barrier generated by the lattice axes in the laboratory frame. Within this angular acceptance, the SF boost is at its peak, but for incidence angles as large as $10 \cdot \Theta_0$ there is a weaker but still non negligible enhancement effect, as it was observed both theoretically~\cite{uggerhoj2005} and experimentally~\cite{soldani2023,pmgthesis}. For reference, the critical angle of the PbWO$_4$ $\langle 001 \rangle$ axis is $\sim$~0.8~mrad, which is much larger than Lindhard's critical angle for the channeling phenomenon for the same axis, i.e., 0.08~mrad at 120~GeV~\cite{soldanithesis}. It is important to observe that, differently from Lindhard's angle, the SF critical angle does not depend on the energy of the incident particle. Details on the derivation of eq.~\eqref{eq:1} can be found elsewhere~\cite{uggerhoj2005}.
	
	\item The energy of the incident particle must be larger than a critical value, usually in the order of tens of GeV~\cite{baier1998_book}. However, if the energy is smaller than the threshold but larger than few GeV, a non negligible enhancement is still observed. For an e$^\pm$ or a photon incident on the PbWO$_4$ $\langle 001 \rangle$ axis the critical value is $\sim$~25~GeV~\cite{soldanithesis}.
\end{enumerate}
The Strong Field effects have been already observed experimentally in multiple studies, which aimed at measuring the enhancement of the BS-PP cross-sections in single element crystals, such as W, Si and Ge~\cite{belkacem1984,moore1996,kirsebom1998,soldani2023}. However, only two studies exist for scintillators in the Strong Field regime, demonstrating the enhancement of the emission of radiation and of the energy deposit as a result of the shower acceleration~\cite{bandiera2018,soldani2024}. Indeed, a few other studies have been done with electrons but at much lower energy, below the SF threshold~\cite{baskov1999}. 

Since the lattice structure does not affect the properties of the hadronic interactions, the difference between a hadronic and an e.m. shower developing inside an oriented crystal should be more pronounced than the ordinary. For this reason, a semi-homogeneous calorimeter featuring one or more oriented layers could represent an optimal choice for performing high-accuracy PID. An oriented detector has never been realized before: there is thus no direct way to quantitatively evaluate the performance increase which could actually be achieved by using the oriented crystals. This is a huge problem, since it is important to know whether it is worth to spend time and money to develop such a detector concept on the large scale required by collider experiments. Fortunately, in the last few years, the ORiEnted calOrimeter (OREO) collaboration has developed several tools for the numerical simulation of the particles interactions inside oriented crystals~\cite{baryshevsky2017,bandiera2019,sytov2023}. These tools were extensively validated by comparing their predictions with the results of multiple beamtest campaigns performed at the CERN PS and SPS~\cite{bandiera2018,soldani2023,soldanithesis,pmgthesis}. Thus, they can be considered as the only instruments currently available to solve this problem and estimate the potential performance gain achievable through the use of oriented crystals.

The aim of this work is thus to evaluate the Particle Identification (PID) potential of a semi-homogeneous eCAL, by means of numerical simulations performed with the tools developed by the OREO collaboration. In particular, we focused on the identification of photons against a neutron background, considering that this is explicitly required for the physics case of the KLEVER experiment, namely the identification of photons produced in $\pi^0$ decays, against a 440~MHz neutron background \cite{moulson2019,moulson2023}.

\section{Materials and methods}
\subsection{Experimental setup (Geant4 simulation)} \label{sec:g4}
The study reported in this article is based on datasets produced using the Geant4 version 11.1 toolkit~\cite{agostinelli2003}. The setup used in the simulation includes a semi-homogeneous eCAL composed of a $5\times 5 \times 4$ matrix of PbWO$_4$ crystals (as shown in figure~\ref{fig:expsetup}). Each crystal has a transverse area of $1\times 1$~cm$^2$ and a thickness of 4~cm. Since the PbWO$_4$ radiation length is $0.89$~cm~\cite{pdg}, each layer is $\sim$~4.5~X$_0$ thick, for a total eCAL length of $\sim$ 18~X$_0$. The simulations were performed in two configurations:
\begin{itemize}
	\item ``Random'' alignment (figure~\ref{fig:expsetup} on the left), meaning that all the eCAL crystals were considered to be made of ordinary, non aligned PbWO$_4$. In this case, the simulations were performed using the \texttt{FTFP\_BERT} Physics List, which implements the standard high-energy e.m. and hadronic processes, including also photo-nuclear interactions~\cite{allison2016}.
	
	\begin{figure}[htb]
		\centering
		\includegraphics[width=0.75\linewidth]{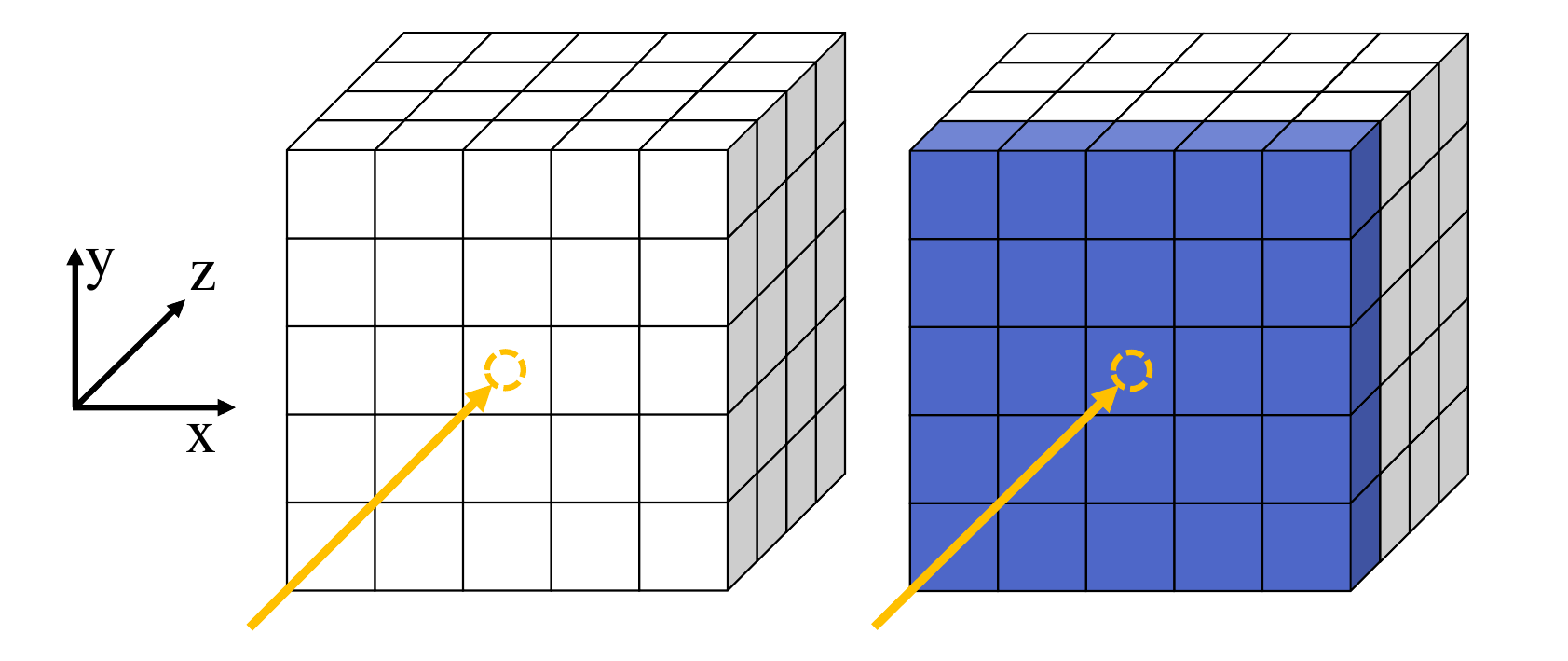}
		\caption{Left: experimental setup used in the Geant4 simulation in ``random'' alignment. Right: setup in ``axial'' configuration. In both figures, each parallelogram represents a PbWO$_4$ crystal either with the $\langle 001 \rangle$ axis aligned with the incident particle beam (dark blue) or randomly aligned (white). The dashed circles (orange) represent the center of the projection of the beam profiles on the face of the calorimeter. (A color version of this figure is available in the online journal)}
		\label{fig:expsetup}
	\end{figure}
	
	\item ``Axial'' alignment (figure~\ref{fig:expsetup} on the right), meaning that all the crystals in the first eCAL layer were considered as oriented along the $\langle 001 \rangle$ axis, while the others were randomly aligned. In this case, the simulations were performed using a modified version of the \texttt{FTFP\_BERT} Physics List, since at present time Geant4 does not implement the physics of the Strong Field regime. In the modified version used for this work, the differential cross-sections of the BS and PP processes were multiplied by a set of coefficients, which increase with the particle energy~\cite{baryshevsky2017,bandiera2019}. These factors were computed beforehand through a full Monte Carlo simulation, where the radiation emission and pair production probability in the axial field of a PbWO$_4$ lattice were computed by directly integrating the quasiclassical Baier-Katkov formula on realistic particle trajectories~\cite{guidi2012,bandiera2015,sytov2019,sytov2023}. This approach has been extensively validated in several studies, by comparing its predictions with the results obtained in beamtest campaigns performed in the last few years by the OREO collaboration~\cite{bandiera2018,soldani2023,soldanithesis,pmgthesis}. Thus, it can be considered as a reliable way to simulate the interactions occurring in an oriented crystal exposed to a high-energy particle beam.
\end{itemize}
In both cases, we did not modify the standard timing cuts implemented by the \texttt{FTFP\_BERT} Physics List, meaning that neutron tracking is stopped after 10~{\textmu}s, while the scoring is performed over an infinite amount of time per each event. It should be mentioned that, by limiting the scoring to a shorter amount of time, it could be possible to achieve a further improvement in the neutron-gamma discrimination performance, since it is known that the hadronic and e.m. showers develop with quite different temporal scales~\cite{benaglia2016,badala2022}. \\

\noindent In both configurations, for each simulated event, the following quantities were recorded:
\begin{itemize}
	\item The class of the incident particle: either a photon (labeled ``positive'' from now onward) or a neutron (labeled ``negative'').
	\item The energy deposited in each crystal.
	\item The energy deposited in each longitudinal layer ($E_{L,i}$, where $i=1,2,3,4$ is the layer index).
	\item The energy deposited in total in the calorimeter ($E_{\text{dep}}$).
\end{itemize}
This means that, for each event, a total of 100 low-level and 5 high-level numerical quantities (from now on, ``features'') were recorded and subsequently analyzed, with the former being the energy deposited in each crystal and the latter the energy deposited in each layer and in the entire eCAL. The addition of these high-level features was considered after a preliminary study phase, where we found that their use led to a slight improvement of the classification performance.

\subsection{Dataset production}
In this work we have studied two different physics scenarios and for each of them we generated one dataset in random configuration and one in axial alignment. The scenarios were the following:
\begin{itemize}
	\item \textbf{Known initial energy}. In this case, the events were generated with an equal amount of neutrons and photons, each one featuring a uniform distribution of the initial energy ($E_{\text{in}}$) in the $26-151$~GeV range, namely the interval where the Geant4 simulation code was experimentally validated and where the SF acceleration is at its peak~\cite{soldani2023,soldani2024}. In particular, this energy range was divided in 2.5~GeV-wide bins: for each bin, 10~000 events per particle type were selected, corresponding to 500~000 photons and 500~000 neutron events in total. In this case, the value of $E_{\text{in}}$ was also recorded on an event-by-event basis and used in the classification process along with the other features listed in the previous paragraph (hence the name of the scenario).
	
	\item \textbf{Known deposited energy}. In this case, a large number of events was generated and then only a part was selected, in order to have an equal amount of neutrons and photons, each one featuring a uniform distribution of the energy deposited in the eCAL ($E_{\text{dep}}$) in the $26-151$~GeV range. In particular, this energy range was divided in 2.5~GeV-wide bins: for each bin, 10~000 events per particle type were selected, corresponding to 500~000 photons and 500~000 neutron events in total. Differently from the first scenario, in this case, the value of $E_{\text{in}}$ was neither recorded nor used in the classification process.
\end{itemize}
The key difference between these scenarios is the fact that we account for the different $n/\gamma$ interaction probabilities only in the first case. In fact, in the second scenario we selected only the events where the incident particles deposited the same energy, independently from their class and from how rare this possibility may be. While the first scenario is the easiest to reproduce experimentally (e.g., in a beamtest, with charged pions instead of neutrons, given the difficulty of producing a pure and monochromatic neutron beam at such high energies), the second is the most interesting for the particle physics experiments. In fact, it represents a real-world scenario, where each event may be classified without knowing the initial energy of the particle generating the signal in the eCAL, but only (at most) its statistical distribution, obtained through Monte Carlo simulations.

In the known $E_{\text{in}}$ scenario we used a $\gamma$ and a neutron beam with a uniform energy spectrum in the $26-151$~GeV range. The beam angular profile was set for both particles to a 2D gaussian with a divergence small enough to guarantee the satisfaction of the Strong Field angular condition ($\sigma_x = \sigma_y = 0.1$~mrad, which is the typical divergence of the electron beams used in the validation of the simulation code). The spatial beam profile was set for both particles to a 2D uniform distribution, covering the entire face of the calorimeter. In the known $E_{\text{dep}}$ scenario we used a $\gamma$ beam with a uniform energy spectrum in the $26-251$~GeV range and a neutron beam with uniform energy spectrum in the $26-501$~GeV range. The angular and spatial beam profiles were defined identically to the first scenario.

\subsection{Particle identification algorithm and metrics} \label{sec:materials_b}
The algorithm chosen to perform the particle identification was the Random Forest (RF) classifier, implemented using the \texttt{scikit-learn} version 1.2 module in Python version 3.11~\cite{pedregosa2011}. This algorithm was chosen due to its relative simplicity and understandability and also due to its relatively good performance in managing large datasets. In fact, these properties have made it widely used in particle and astroparticle physics~\cite{graczykowski2022,albert2008}.

The RF classifier was separately optimized for each of the four analyzed datasets (one in random and one in axial for each of the two scenarios). The aim of the optimization was to determine the combination of hyperparameters which maximized the classification performance: to achieve this purpose, a brute-force Grid Search was performed, scanning over the hyperparameter space reported in table~\ref{tab:GS}. For each combination of the hyperparameters, a 5-Fold Cross-validation was performed, meaning that the RF was trained and tested 5 times using each time the same dataset splitted in a different way, with 80\% of the available data used for the training (Training Set, TrS) and the remaining 20\% for the testing (Test Set, TeS). 

\begin{table}[ht]
	\centering
	\caption{\label{tab:GS} List of the hyperparameters considered in the optimization of the Random Forest classifier. The optimization was repeated twice for each physics scenario (each time, one in random and one in axial) and it always converged to the combination shown on the right. The hyperparameter space analyzed in the optimization was composed by considering every possible configuration of each of these hyperparameters, for a total of 36 cases per dataset. All the hyperparameters not mentioned in this table were fixed to their default values~\cite{sklearnrfweb}.}
	\smallskip
	\begin{tabular}{| c c c |}
		\hline
		\emph{hyperparameter} & \emph{Tested values} & \emph{Optimal value} \\
		\hline
		\texttt{n\_estimators} 	& 50, 100, 150, 200 	& 200 \\
		\texttt{depth} 			& `None', 5, 15 		& `None' \\
		\texttt{max\_features} 	& `sqrt', 5, 15 		& 15 \\
		\hline
	\end{tabular}
\end{table}

\noindent During the training phase, the data were pre-processed as follows:
\begin{enumerate}
	\item The features were normalized to a null average and unit variance.
	
	\item A Principal Component Analysis was performed. The Principal Components (PCs) were then sorted by decreasing Explained Variance (EVs).
	
	\item To reduce the dimensionality of the dataset, only the Principal Components whose cumulative EV amounted to 95\% of the total were used in the analysis. On average, only half of the PCs were required to reach this threshold, with a small dependence on the physics scenario and the crystal alignment (table~\ref{tab:PCA}).
	
	\begin{table}[ht]
		\centering
		\caption{\label{tab:PCA} Average number of Principal Components necessary to reach 95\% of the total cumulative EV, in both of the physics scenarios (known $E_{\text{in}}$ and $E_{\text{dep}}$) and the crystal alignment conditions (axial and random). In both cases, the number of initially available PCs was 105.}
		\smallskip
		\begin{tabular}{| c c c |}
			\hline
			\emph{Scenario} & \emph{Crystal orientation} & \emph{Number of PCs} \\
			\hline
			Known $E_{\text{in}}$ 		& Random 	& 52 \\
			& Axial 	& 50 \\
			Known $E_{\text{dep}}$ 		& Random 	& 50 \\
			& Axial 	& 45 \\
			\hline
		\end{tabular}
		
	\end{table}
\end{enumerate}
After the pre-processing, the RF parameters were learned using the TrS. Then, the TeS was pre-processed with the same scaling and dimensionality reduction techniques used in the training. The hyperparameter configuration which determined the highest average accuracy was considered as the optimal point of work, with the accuracy being defined as the fraction of correct predictions over the total number of events, and the average being made over the 5 values obtained in the Cross-validation. The optimal hyperparameter configuration has always been found to be the one which led to the most complex RF classifier (table~\ref{tab:GS}), i.e., the one with the largest number of trees in the Forest, each one featuring the most complex structure possible. In principle, by increasing further the number of trees in the Forest or the number of features considered at each split, it should be possible to achieve an even better performance. However, during the optimization phase, it was noted that the 5-Fold classification accuracy had reached a plateau while increasing either \texttt{n\_estimators} or \texttt{max\_features} beyond the maximum values tested. In fact, by further increasing these parameters, the accuracy was found to grow by not more than $\sim$~0.25\%, while the computation time required for the training of the RF classifier grew exponentially. For this reason and also in order to avoid over-fitting, it was chosen to not perform a wider grid search.

After the optimization, the 5-Fold Cross-validation was repeated separately for each dataset and scenario, using the optimal hyperparameters configuration. This time, the Test Sets were divided in subsets depending on the value of either $E_{\text{in}}$ or $E_{\text{dep}}$ (respectively for the known $E_{\text{in}}$ and $E_{\text{dep}}$ scenarios). Afterwards, the value of each classification score was computed separately for each subset and averaged over the 5 values obtained in the Cross-validation. Moreover, the Receiver Operating Characteristic (ROC) curves and the areas subtended by them (i.e., the Areas Under the Curve, AUCs) were computed and used as an additional estimators of the classifier performance.

\section{Results and discussion} \label{sec:results}
\subsection{Known $E_{\text{in}}$ scenario}
\begin{figure}[htb]
	\centering
	\includegraphics[width=0.75\linewidth]{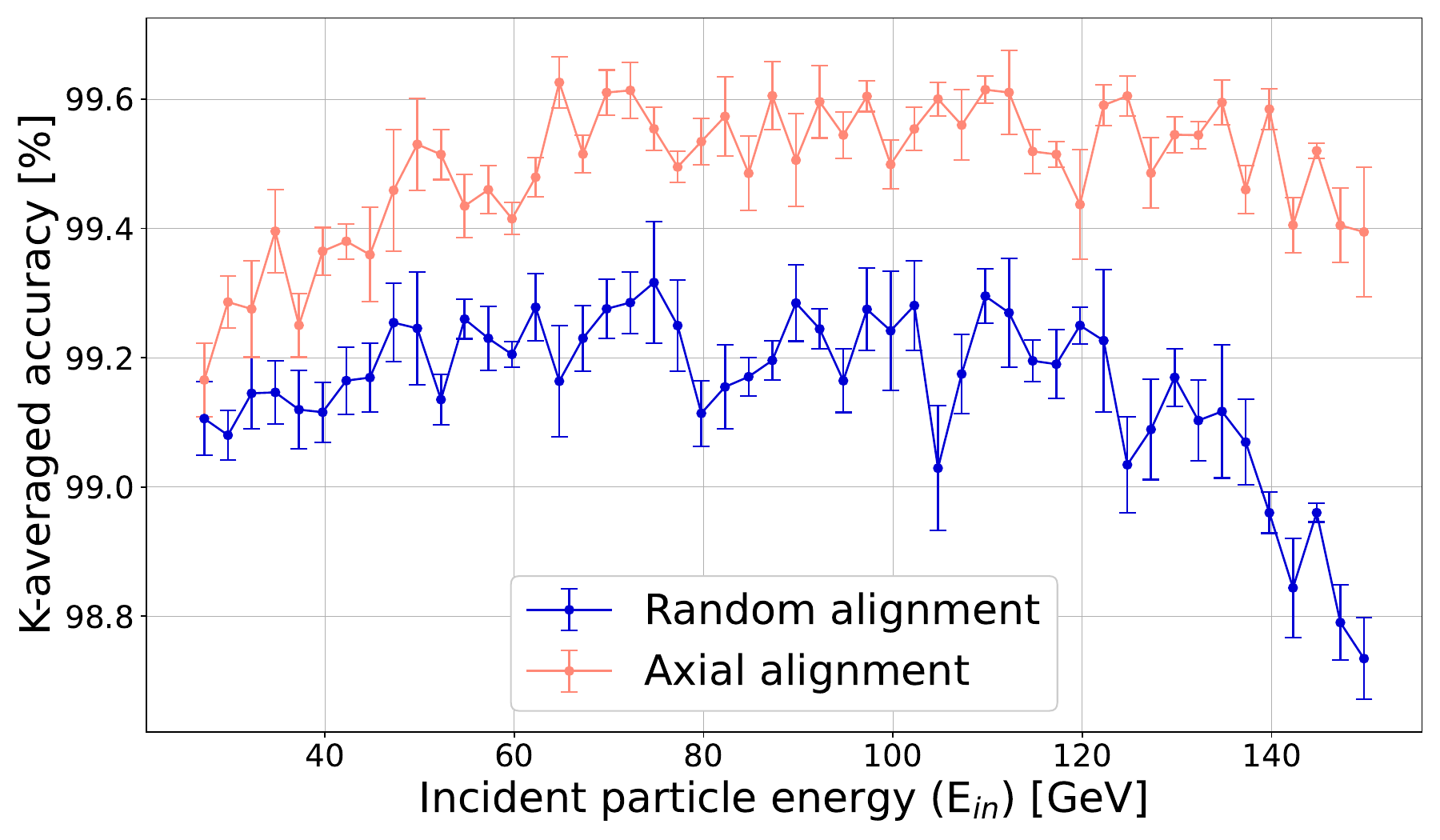}
	\caption{Dependence of the 5-Fold averaged classification accuracy on the energy of the incident particles, obtained in the known $E_{\text{in}}$ scenario. The errors were computed as standard deviations of the average accuracy. (A color version of this figure is available in the online journal)}
	\label{fig:EIN_rf_results_allacc}
\end{figure}

\begin{figure}[htb]
	\centering
	\includegraphics[width=0.75\linewidth]{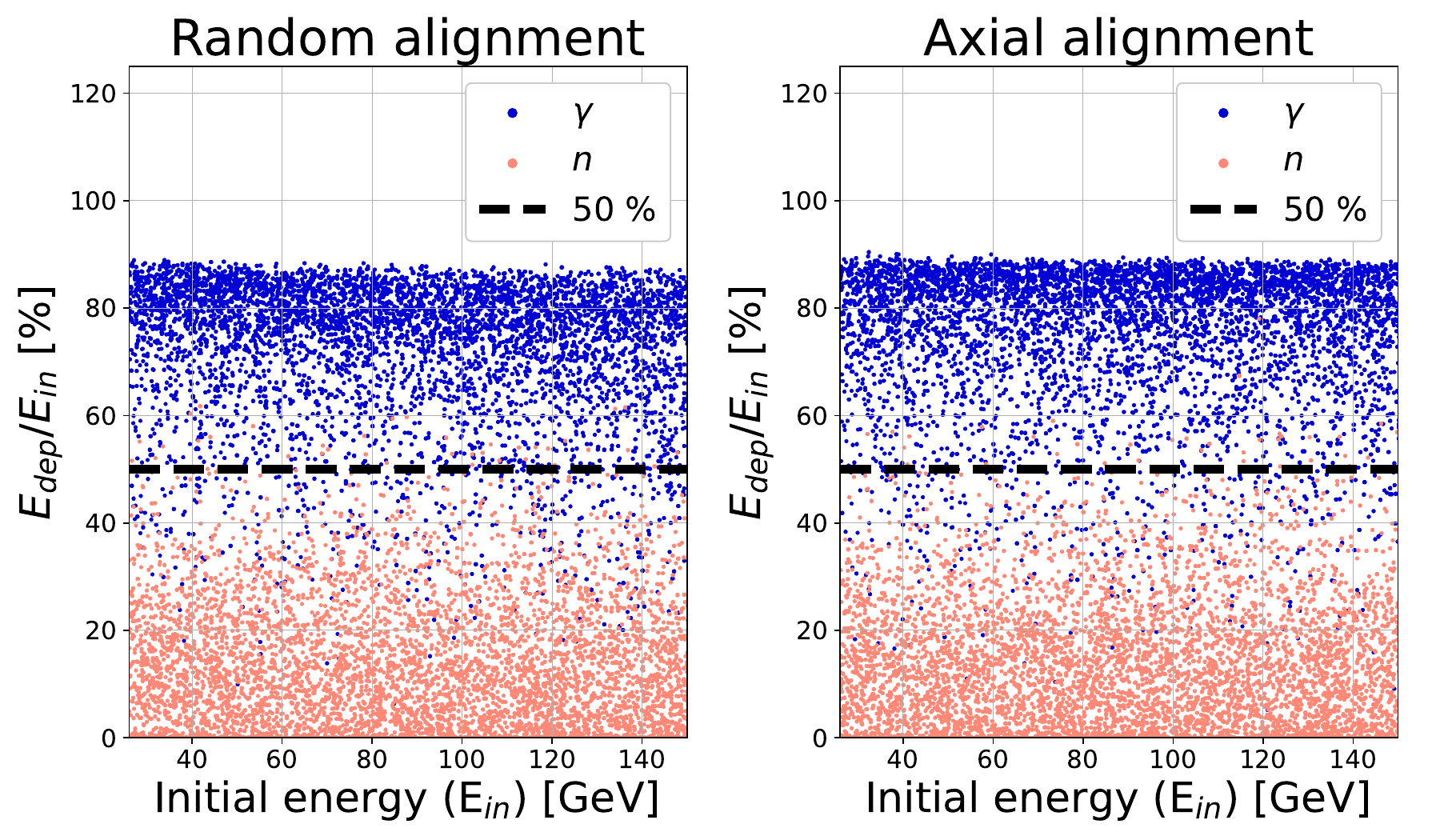}
	\caption{Correlation between the initial energy of the photons/neutrons (blue/light red) and the fraction of the particles initial energy deposited in the calorimeter (i.e., $E_{\text{dep}}/E_{\text{in}}$), obtained in the known $E_{\text{in}}$ scenario. The black dashed line shows where $E_{\text{dep}}$ corresponds to 50\% of $E_{\text{in}}$. Only 10~000 events per configuration per particle type are shown in the scatter plots, instead of the full statistics used in the study, for better visual clarity. (A color version of this figure is available in the online journal)}
	\label{fig:EIN_scatter_EdepvsEin}
\end{figure}

\begin{figure}[htb]
	\centering
	\includegraphics[width=0.75\linewidth]{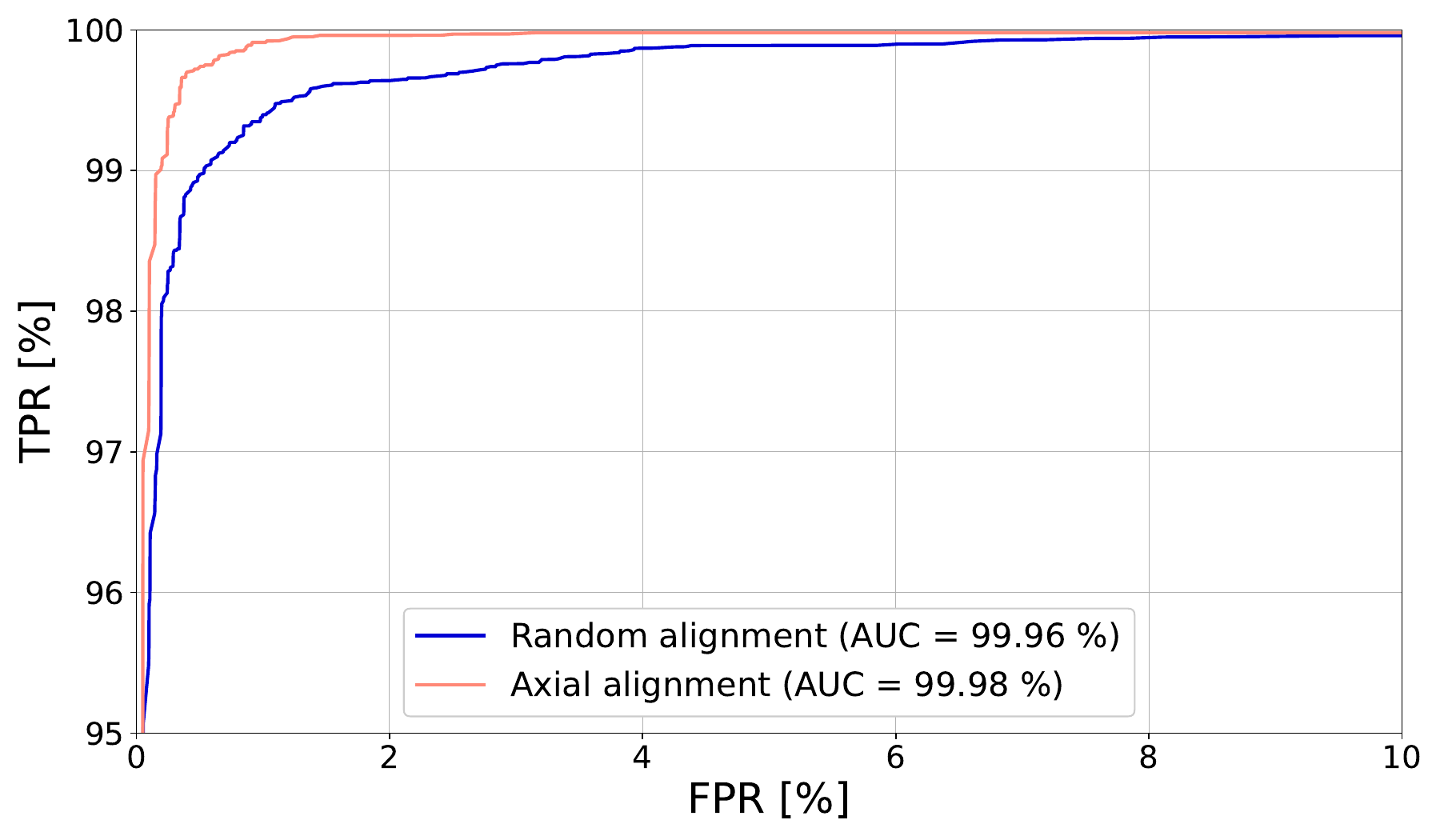}
	\caption{Example ROC curve, obtained in the known $E_{\text{in}}$ scenario by analyzing only the events where the neutrons and photons featured an initial energy of ($120 \pm 1.25$)~GeV. In this plot and in the following, FPR is the False Positive Rate and TPR is the True Positive Rate. (A color version of this figure is available in the online journal)}
	\label{fig:EIN_rf_results_ROCcurves_120GeV}
\end{figure}

Figure~\ref{fig:EIN_rf_results_allacc} shows how the RF classification accuracy depends on the initial energy of the particles, in the known $E_{\text{in}}$ scenario. It can be seen that the accuracy is essentially constant and better than 99\% in both the random and axial configuration, with a small dependence on the crystal orientation. These values reflect the fact that photons and neutrons deposit their energy through showers featuring quite different spatial profiles. In fact, as it can be seen in figure~\ref{fig:EIN_scatter_EdepvsEin}, photons deposit in the active volume usually more than 70\% of their initial energy, while the remaining part is lost primarily due to longitudinal and transverse leakage. On the other hand, neutrons deposit very rarely more than $10-30\%$ of their energy in the calorimeter volume, since the PbWO$_4$ neutron interaction length ($\lambda_{\text{int}}$) is 20.27~cm~\cite{pdg}: the entire eCAL is only $0.79 \, \lambda_{\text{int}}$ thick and thus it never fully contains the hadronic showers. From these figures we can deduce that the discrimination is completely dominated by the knowledge of the initial and deposited energy, while the more detailed informations obtained through the eCAL segmentation are significantly less important. In fact, if the discrimination was simply carried out by considering as photons all the events where $E_{\text{dep}} / E_{\text{in}} > 50\%$ and as neutrons all the others, an accuracy of 95.68\% (96.30\%) would have been found in the random (axial) configuration. Similar considerations can be deduced by analyzing the ROC curves of the RF classifier, an example of which is shown in figure~\ref{fig:EIN_rf_results_ROCcurves_120GeV}: the True Positive Rate (TPR) is almost always equal to the unity and the AUCs are very close to 100\%.

\begin{figure}[htb]
	\centering
	\includegraphics[width=0.75\linewidth]{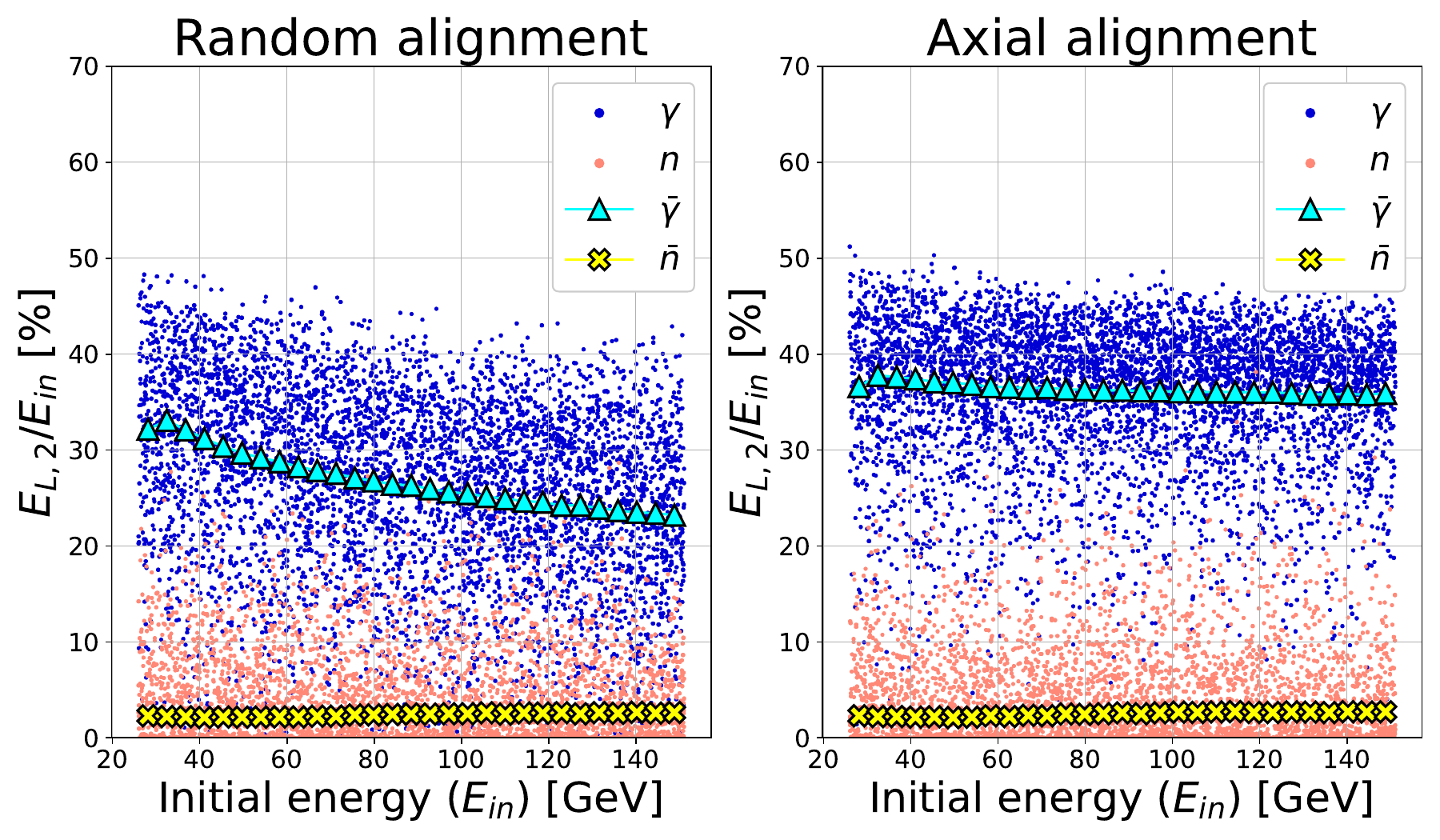}
	\caption{Correlation between the initial energy of the photons/neutrons (blue/light red) and the fraction of the particles initial energy deposited in the second eCAL layer, obtained in the known $E_{\text{in}}$ scenario. The markers represent the average energy deposited in the second eCAL layer by the incident photons/neutrons (cyan triangles/yellow crosses). The average values were computed using the full statistics used in the study, while only 10~000 events per dataset were shown, for better visual clarity. The difference between the average energy deposited by the incident photons in random configuration and axial alignment is due to the e.m. shower acceleration induced in the oriented crystalline layer. (A color version of this figure is available in the online journal)}
	\label{fig:EIN_scatter_EL1vsEin}
\end{figure}

In this scenario it is possible to appreciate how much the e.m. shower is accelerated when the Strong Field condition is satisfied. In order to do so, it is sufficient to measure the fraction of the particles initial energy deposited in the second eCAL layer ($E_{L,2}/E_{\text{in}}$), corresponding to the point in space where the photon-initiated showers usually reach their peak~\cite{gianotti2003}. As shown in figure~\ref{fig:EIN_scatter_EL1vsEin}, we found that, on average, $E_{L,2}/E_{\text{in}}$ presents a decreasing trend for photons incident on the randomly aligned calorimeter, ranging from $\sim$~35\% to $\sim$~25\%, while the average ratio measured for photons incident on the axially aligned detector is almost constant and equal to $\sim$~35\%. In both cases, neutrons deposit on average $\sim$~2\% of their energy in the same layer. This observation suggests that even in a calorimeter thinner than the one studied in this work, the presence of one or more oriented layers could be useful for achieving high-accuracy particle identification.

\subsection{Known $E_{\text{dep}}$ scenario}
\begin{figure}[htb]
	\centering
	\includegraphics[width=0.75\linewidth]{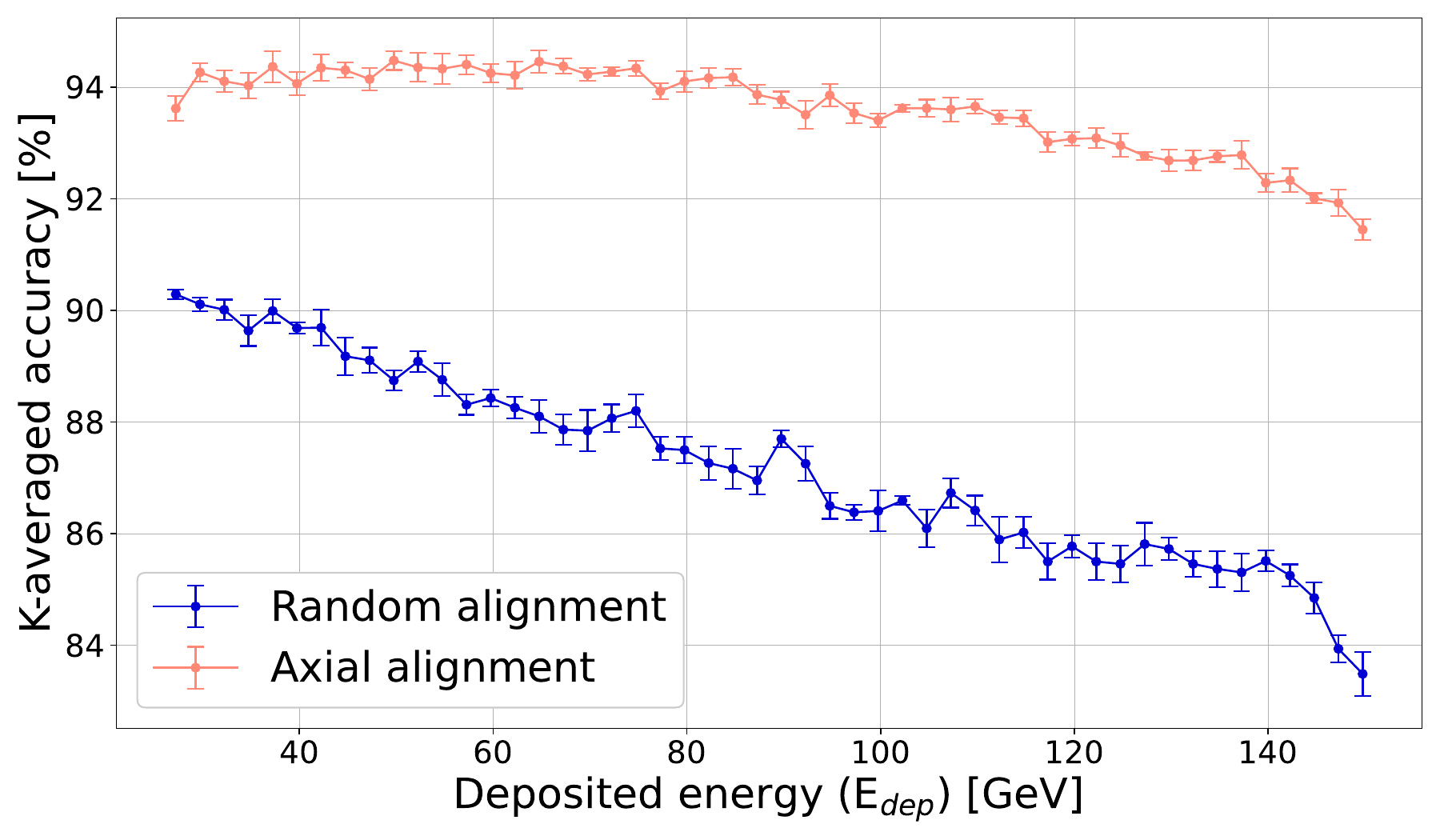}
	\caption{Dependence of the 5-Fold averaged classification accuracy on the energy deposited by the incident particles in the calorimeter, obtained in the known $E_{\text{dep}}$ scenario. The errors were computed as standard deviations of the average accuracy. (A color version of this figure is available in the online journal)}
	\label{fig:EDEP_rf_results_allacc}
\end{figure}

Figure~\ref{fig:EDEP_rf_results_allacc} shows how the RF classification accuracy depends on the total energy deposited in the eCAL, in the known $E_{\text{dep}}$ scenario. Here, a much larger difference between the axial and random configuration can be observed: in the random case, the accuracy decreases monotonically with $E_{\text{dep}}$, while in the axial case it is constant up to $\sim 100$~GeV and then it decreases slowly. The curve obtained in the random case is explained by considering that relativistic hadrons leave an increasing fraction of their energy in the form of e.m. showers~\cite{gianotti2003}. In this way, their energy deposit profiles become intrinsically similar to those induced by the photons and thus the accuracy naturally decreases. On the other hand, in the axial configuration, the same effect occurs, but it is canceled out by the increasing intensity of the Strong Field enhancement of the BS and PP cross-sections: the resulting accuracy is larger than the one obtained in the random configuration by $5-8\%$. At energies larger than 100~GeV, the similarity between photons and neutrons increases to the point where the accuracy decreases because the e.m. part of the hadronic shower can develop already in the first radiation lengths, where it is accelerated. It is important to observe that this does not imply that the Strong Field is directly affecting the hadronic interactions. On the contrary, this is a consequence of the fact that the datasets here analyzed were constructed by selecting only the rare events where the incident neutrons deposited in the eCAL a large fraction of their energy. Beyond $\sim$100~GeV, this effect happens mostly if a hadron gives rise to an e.m. shower already in the first eCAL layer: it is only the e.m. component of the shower whose development is then accelerated, if it falls within the SF acceptance angular range. It should be noted that this may not always be the case, since hadron-generated e.m. showers are mainly induced by $\pi^0$ decays and nuclear de-excitations, which are processes with a relatively wide angular divergence.

\begin{figure}[htb]
	\centering
	\includegraphics[width=0.75\linewidth]{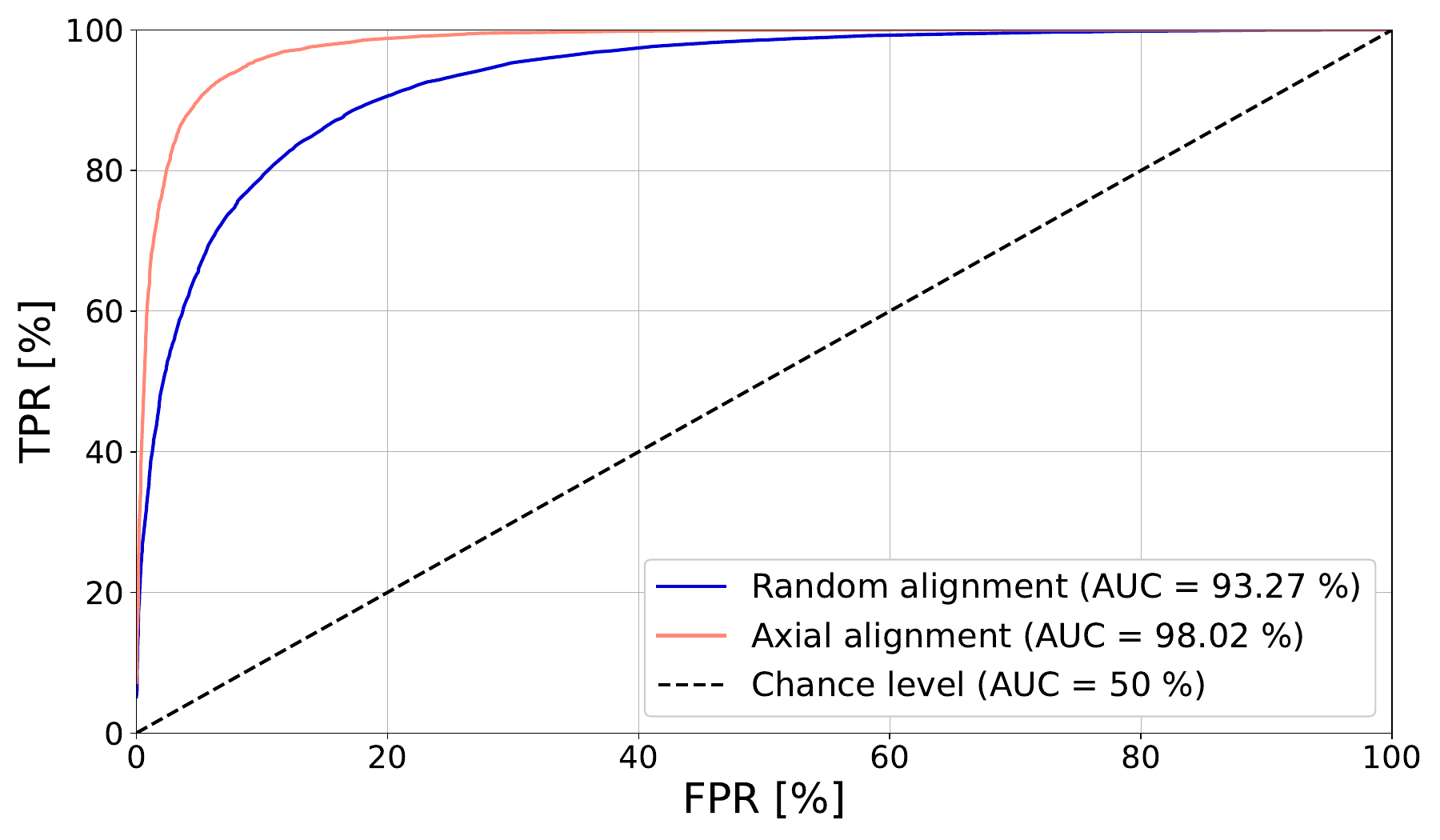}
	\caption{Example ROC curve, obtained in the known $E_{\text{dep}}$ scenario by analyzing only the events where the neutrons and photons deposited an energy of ($120 \pm 1.25$)~GeV in the calorimeter. The black dashed line shows the ROC curve of a pure-chance classifier (i.e., the one with a 50\% AUC). (A color version of this figure is available in the online journal)}
	\label{fig:EDEP_rf_results_ROCcurves_120GeV}
\end{figure}

\begin{figure}[htb]
	\centering
	\includegraphics[width=0.75\linewidth]{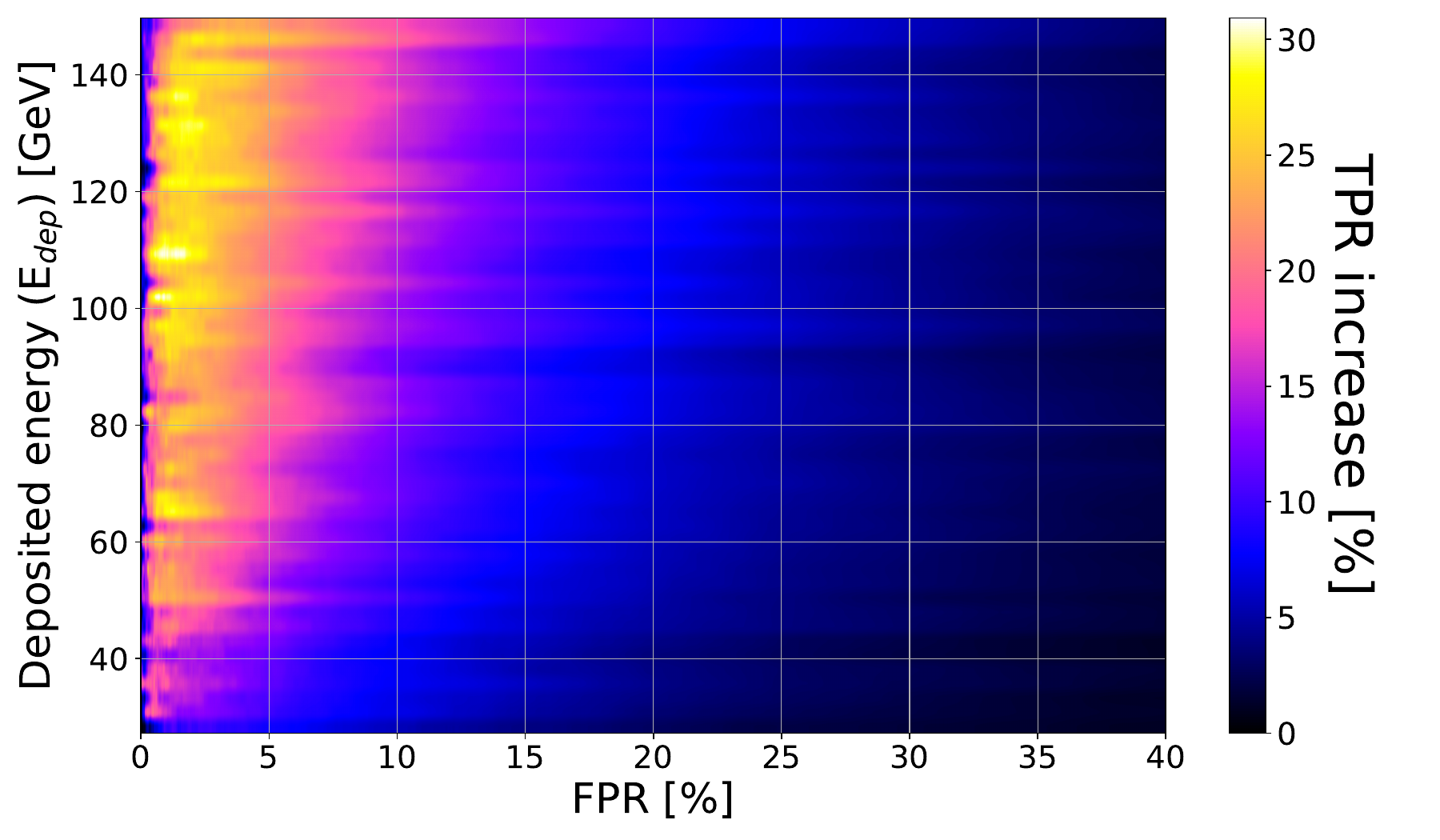}
	\caption{2D histogram showing the difference between the ROC curves measured in the axial and random configuration, as a function of the FPR and of the energy deposited in the calorimeter ($E_\text{dep}$). It can be observed that the largest TPR increase is achieved in the region where the FPR is minimum, which is generally the point where the classifier should be operated, in order to maximize the background rejection. (A color version of this figure is available in the online journal)}
	\label{fig:EDEP_rf_results_ROCcurves_increases2D}
\end{figure}

Figure~\ref{fig:EDEP_rf_results_ROCcurves_120GeV} shows an example ROC curve, obtained by analyzing only the events with a deposited energy of ($120~\pm~1.25$)~GeV. The smaller AUCs here obtained confirm that in this scenario the PID process is effectively more error prone. Moreover, we can appreciate once again how the oriented crystals improve the performance of the RF classifier: the TPR measured in the axial case is larger than the one obtained in the random configuration by $\sim$~15\% (when the FPR is $\sim$ 5\%). This means that if the RF is used with thresholds which minimize the fraction of misidentified neutrons (FPR $\lesssim 10\%$), it is possible to significantly improve the classification efficiency, if the first eCAL layer is aligned. The dependence of this efficiency enhancement on the FPR and the deposited energy is shown in figure~\ref{fig:EDEP_rf_results_ROCcurves_increases2D}, where it can be seen that increases as large as 30\% can be reached, in the lowest-FPR region. However, when determining the specific thresholds to be used, the trade-off between low FPR and high TPR values must always be carefully balanced with respect to the goals of the experiment in which this calorimeter is used.

\begin{figure}[htb]
	\centering
	\includegraphics[width=0.75\linewidth]{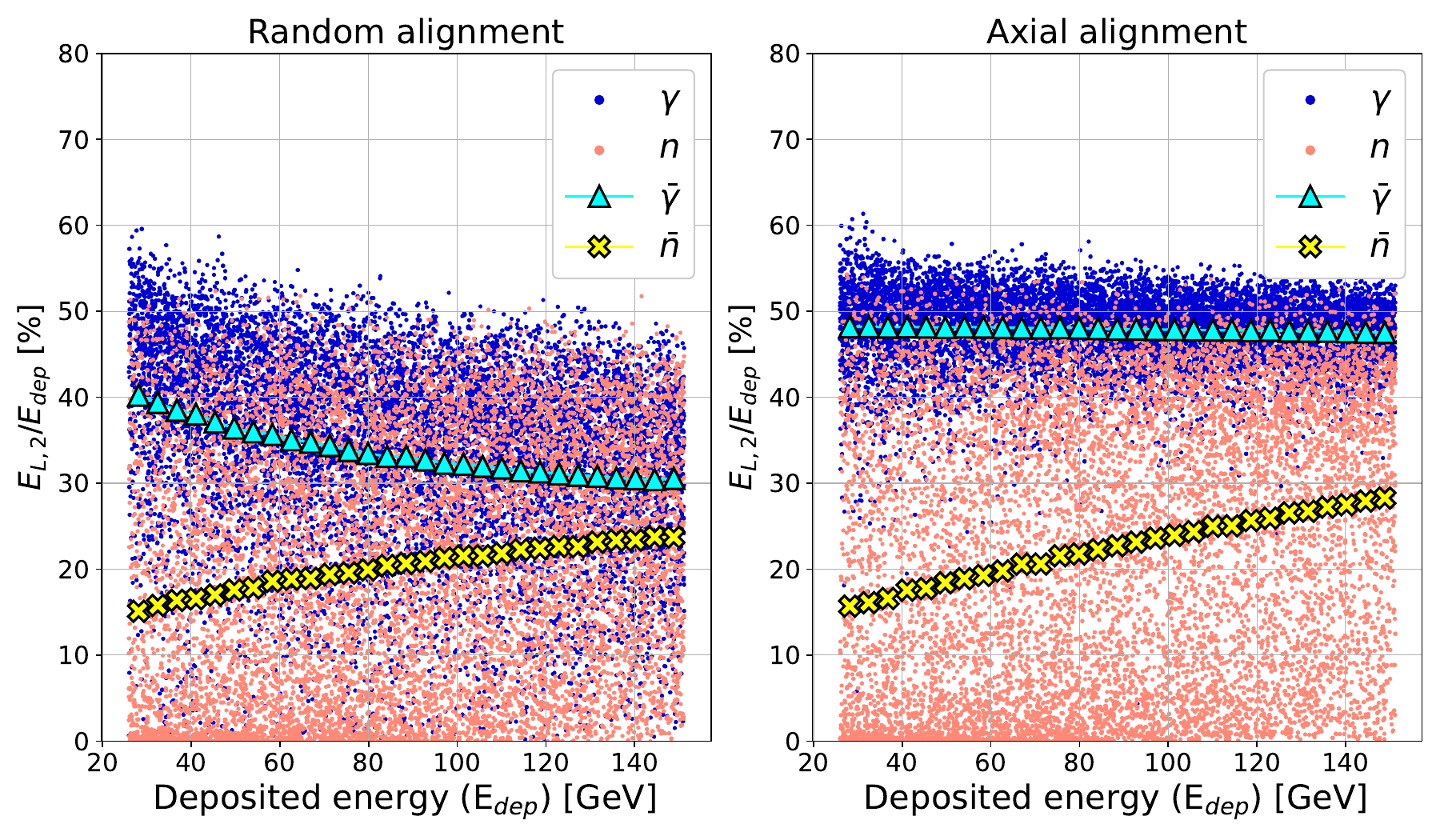}
	\caption{Correlation between the energy deposited in the calorimeter by the photons/neutrons (blue/light red) and the fraction of the latter deposited in the second eCAL layer, obtained in the known $E_{\text{dep}}$ scenario. The markers represent the average energy deposited in the second eCAL layer by the incident photons/neutrons (cyan triangles/yellow crosses). The average values were computed using the full statistics used in this study, while only 10~000 events per dataset were shown, for better visual clarity. The difference between the average energy deposited by the incident photons in random and axial configuration is due to the e.m. shower acceleration induced by the oriented crystalline layer. (A color version of this figure is available in the online journal)}
	\label{fig:EDEP_scatter_EL1vsEdep}
\end{figure}

As in the previous scenario, we can also qualitatively study the separation between the photon and neutron events in the $E_{L,2}/E_{\text{dep}}$ vs $E_{\text{dep}}$ plane (figure~\ref{fig:EDEP_scatter_EL1vsEdep}). We can see that there is no longer a clear separation between the two classes and that their overlap is significantly worse in the random case. In fact, the average $E_{L,2}/E_{\text{dep}}$ ratio decreases from $\sim$~50\% to $\sim$~30\% for the photons in the random case, while it remains fixed around $\sim$~50\% in the axial configuration. Instead, in both configurations, the average neutron ratio grows from 10\% to 25\%, thus showing a much better separation from the photon curve in the axial case.

\section{Conclusions and outlook} \label{sec:conclusions}
In this work we have studied the particle identification capability of a semi-homogeneous e.m. calorimeter, with fine transverse and longitudinal segmentation. We have focused on identifying photons from a neutron background, using for the discrimination a Random Forest algorithm trained on the energies deposited in all of the eCAL crystals. We have demonstrated that the use of oriented crystals in the first eCAL layer significantly improves the classification accuracy and reduces the fraction of misidentified photons. Such a calorimeter could be essential in many high-intensity environments, where the traditional $n/\gamma$ classification techniques may not be as efficient as normal. Such environments include, for instance, the third phase of the HIKE project and the instrumented beamline for a possible future Muon Collider. Other applications may also include source-pointing satellite-born $\gamma-$ray telescopes, similar to the currently operating Fermi Large Area Telescope (LAT), an instrument used for the observation of the $\gamma-$ray sky from $\sim$~50~MeV up to $\sim$~2~TeV, which includes on board an 8.6~X$_0$ thick CsI(Tl) calorimeter~\cite{atwood2009}. In such cases, the detectors on board of the mission must be able to discriminate the incident photons from a huge cosmic ray background, primarily composed of high-energy protons. Since in a satellite-born telescope the weight of the calorimeter directly determines the cost of the mission, the use of one or more oriented-crystal layers could lead to a significant improvement of its PID capability at no additional cost. However, given the small angular acceptance of the Strong Field regime, such an improvement would only be achievable in the center of the satellite Field Of View (FOV), while in the remaining part the detector would continue to work normally. As a consequence, such a satellite detector would achieve the best performance in source-pointing mode. \\

By following the same approach used in this study it is possible to qualify the PID potential of any crystalline calorimeter, independently from its layout and from the features of the incident particle beams. On the other hand, it is non trivial to find the optimal oriented calorimeter design, i.e., the best combination of the total eCAL thickness, the segmentation step and the extent of the oriented region, since they can affect the identification performance in different and counterintuitive ways. As we have seen, the currently implemented model of e.m. shower formation in oriented crystals is simple and powerful, but it does not include all the features of coherent interaction of particles with crystals and, especially, it only works above $\sim$~25 GeV, while it is known that orientational effects exist also at the GeV scale. Indeed, in one of the next versions of Geant4 the full model of radiation emission described in \cite{sytov2023} (which is based on a different code than the one used for this study) will be implemented, while the full pair production model is still under development. Such models will be even more accurate than the one used in this work and will be a fundamental tool for predicting the interactions occurring in oriented crystals even at the microscopic level. From the experimental point of view, the results of the ongoing OREO project, which aims at developing the first small-scale prototype of an oriented calorimeter composed of oriented PbWO$_4$ crystals \cite{bandiera2023}, will be crucial to probe the model and to collect important information for calorimeter construction, such as the Molière radius of the oriented crystals and other relevant physical quantities. In conclusion, this work represents only the first (but necessary) step towards PID with an oriented calorimeter and it confirms the intuition that this is a promising and novel approach in that direction.

\section*{Acknowledgements}
This work was supported by INFN CSN/5 (OREO and GEANT4INFN) and CSN/1 (NA62 experiment, RD-FLAVOUR project). This work was also partially supported by the European Commission, through the H2020-INFRAINNOV AIDAINNOVA (G.A. 101004761), H2020-MSCA-RISE N-LIGHT (G.A. 872196), H2020-MSCA-IF-Global TRILLION (G.A. 101032975) and the HorizonEU EIC-PATHFINDER-OPEN TECHNO-CLS (G.A. 101046458) projects. We acknowledge financial support under the National Recovery and Resilience Plan (NRRP), Call for tender No. 104 published on 02.02.2022 by the Italian Ministry of University and Research (MUR), funded by the European Union – NextGenerationEU – Project Title : "Intense positron source Based On Oriented crySTals - e+BOOST" 2022Y87K7X– CUP I53D23001510006. The authors wish to thank the CERN LXPLUS and Cloud Infrastructure services and their staff for providing the high performance computing resources used in the execution of this work and for their support. The authors wish to thank M. Landoni for fruitful discussions on the particle identification techniques used in this work.

\clearpage
\bibliography{Bibliography_file.bib}

\providecommand{\href}[2]{#2}\begingroup\raggedright\begin{thebibliography}{10}

\bibitem{gianotti2003}
C.W.~Fabjan and F.~Gianotti, \emph{Calorimetry for particle physics},
  \href{https://doi.org/10.1103/RevModPhys.75.1243}{\emph{Rev. Mod. Phys.}
  {\bfseries 75} (2003) 1243}.

\bibitem{paesani2022}
D.~Paesani et~al., \emph{{Mechanical Design of an Electromagnetic Calorimeter
  Prototype for a Future Muon Collider}},
  \href{https://doi.org/10.3390/instruments6040063}{\emph{Instruments}
  {\bfseries 6} (2022) }.

\bibitem{cantone2023}
C.~Cantone et~al., \emph{{Beam test, simulation, and performance evaluation of
  PbF2 and PWO-UF crystals with SiPM readout for a semi-homogeneous calorimeter
  prototype with longitudinal segmentation}},
  \href{https://doi.org/10.3389/fphy.2023.1223183}{\emph{Front. Phys.}
  {\bfseries 11} (2023) }.

\bibitem{ceravolo2022}
S.~Ceravolo et~al., \emph{{CRILIN: A CRystal calorImeter with Longitudinal
  InformatioN for a future Muon Collider}},
  \href{https://doi.org/10.1088/1748-0221/17/09/p09033}{\emph{J. Instrum.}
  {\bfseries 17} (2022) }.

\bibitem{ceravolo2022tris}
S.~Ceravolo et~al., \emph{{CRILIN: A Semi-Homogeneous Calorimeter for a Future
  Muon Collider}},
  \href{https://doi.org/10.3390/instruments6040062}{\emph{Instruments}
  {\bfseries 6} (2022) }.

\bibitem{moulson2019}
M.~Moulson, \emph{{KLEVER: An experiment to measure BR($K_L \to \pi^0 \nu
  \bar{\nu}$) at the CERN SPS}},  in \emph{{International Conference on Kaon
  Physics}}, vol.~1526, p.~012028, 2019,
  \href{https://doi.org/10.1088/1742-6596/1526/1/012028}{DOI}.

\bibitem{moulson2023}
{\scshape {HIKE}} collaboration, \emph{{HIKE: High Intensity Kaon Experiments
  at the CERN SPS}},
  \href{https://doi.org/10.1088/1742-6596/2446/1/012036}{\emph{Journal of
  Physics: Conference Series} {\bfseries 2446} (2023) 012036}.

\bibitem{gil2022}
{The HIKE collaboration}, ``{HIKE, High Intensity Kaon Experiments at the CERN
  SPS: Letter of Intent}.'' CERN-SPSC-2022-031, SPSC-I-257,
  \url{https://cds.cern.ch/record/2839661}, 2022.

\bibitem{hike2023}
{The HIKE Collaboration}, \emph{{High Intensity Kaon Experiments (HIKE) at the
  CERN SPS: Proposal for Phases 1 and 2}},  2023.
\newblock 10.48550/arXiv.2311.08231.

\bibitem{benaglia2016}
A.~Benaglia et~al., \emph{{Space-Time Development of Electromagnetic and
  Hadronic Showers and Perspectives for Novel Calorimetric Techniques}},
  \href{https://doi.org/10.1109/TNS.2016.2527758}{\emph{{IEEE Trans. Nucl.
  Sci.}} {\bfseries 63} (2016) 574}.

\bibitem{badala2022}
A.~Badala et~al., \emph{{Trends in particle and nuclei identification
  techniques in nuclear physics experiments}},
  \href{https://doi.org/10.1007/s40766-021-00028-5}{\emph{{Riv. del Nuovo
  Cim.}} {\bfseries 45} (2022) 189}.

\bibitem{kumakhov1977}
M.A.~Kumakhov, \emph{Theory of radiation of charged particles channeled in a
  crystal}, \href{https://doi.org/10.1002/pssb.2220840106}{\emph{{Phys. Status
  Solidi B}} {\bfseries 84} (1977) 41}.

\bibitem{uggerhoj2005}
U.I.~Uggerh\o{}j, \emph{{The interaction of relativistic particles with strong
  crystalline fields}},
  \href{https://doi.org/10.1103/RevModPhys.77.1131}{\emph{Rev. Mod. Phys.}
  {\bfseries 77} (2005) 1131}.

\bibitem{sorensen1987_nature}
A.H.~S\o{}rensen and E.~Uggerh\o{}j, \emph{Channelling and channelling
  radiation}, \href{https://doi.org/10.1038/325311a0}{\emph{Nature} {\bfseries
  325} (1987) }.

\bibitem{sorensen1996_notes}
A.H.~S\o{}rensen, \emph{Channeling, bremsstrahlung and pair creation in single
  crystals}, \href{https://doi.org/10.1016/0168-583X(96)00349-7}{\emph{Nucl.
  Instrum. Methods Phys. Res. B} {\bfseries 119} (1996) }.

\bibitem{baryshevski1983}
V.G.~Baryshevsky and V.V.~Tikhomirov, \emph{Creation of transversely polarized
  high-energy electrons and positrons in crystals}, {\emph{Sov. Phys. JETP}
  {\bfseries 58} (1983) }.

\bibitem{baier1998_book}
V.N.~Baier et~al., \emph{Electromagnetic Processes at High Energies in Oriented
  Single Crystals}, World Scientific (1998),
  \href{https://doi.org/10.1142/2216}{10.1142/2216}.

\bibitem{kimball1984a}
J.C.~Kimball and N.~Cue, \emph{Constant field approximation for the
  crystal-assisted pair-creation process},
  \href{https://doi.org/10.1016/0168-583X(84)90147-2}{\emph{Nucl. Instrum.
  Methods Phys. Res. B} {\bfseries 2} (1984) }.

\bibitem{baryshevski1989}
V.G.~Baryshevsky and V.V.~Tikhomirov, \emph{Synchrotron-type radiation
  processes in crystals and polarization phenomena accompanying them},
  \href{https://doi.org/10.1070/PU1989v032n11ABEH002778}{\emph{Sov. Phys.
  Uspekhi} {\bfseries 32} (1989) 1013}.

\bibitem{soldani2024}
M.~Soldani et~al., \emph{Acceleration of electromagnetic shower development and
  enhancement of light yield in oriented scintillating crystals}, .

\bibitem{bandiera2018}
L.~Bandiera et~al., \emph{Strong reduction of the effective radiation length in
  an axially oriented scintillator crystal},
  \href{https://doi.org/10.1103/PhysRevLett.121.021603}{\emph{Phys. Rev. Lett.}
  {\bfseries 121} (2018) }.

\bibitem{soldanithesis}
M.~Soldani, \emph{{Innovative applications of strong crystalline field effects
  to particle accelerators and detectors}}, Ph.D. thesis, Ferrara University,
  2023.

\bibitem{belkacem1984}
A.~Belkacem et~al., \emph{{Observation of Enhanced Pair Creation for 50--110
  GeV Photons in an Aligned Ge Crystal}},
  \href{https://doi.org/10.1103/PhysRevLett.53.2371}{\emph{Phys. Rev. Lett.}
  {\bfseries 53} (1984) 2371}.

\bibitem{moore1996}
R.~Moore et~al., \emph{{Measurement of pair-production by high energy photons
  in an aligned tungsten crystal}},
  \href{https://doi.org/10.1016/0168-583X(96)00347-3}{\emph{Nucl. Instrum.
  Methods Phys. Res. B} {\bfseries 119} (1996) 149}.

\bibitem{kirsebom1998}
K.~Kirsebom et~al., \emph{{Pair production by 5--150 GeV photons in the strong
  crystalline fields of germanium, tungsten and iridium}},
  \href{https://doi.org/https://doi.org/10.1016/S0168-583X(97)00589-2}{\emph{Nucl.
  Instrum. Methods Phys. Res. B} {\bfseries 135} (1998) 143}.

\bibitem{soldani2023}
M.~Soldani et~al., \emph{{Strong enhancement of electromagnetic shower
  development induced by high-energy photons in a thick oriented tungsten
  crystal}}, \href{https://doi.org/10.1140/epjc/s10052-023-11247-x}{\emph{Eur.
  Phys. J. C} {\bfseries 83} (2023) }.

\bibitem{baskov1999}
V.A.~Baskov et~al., \emph{Electromagnetic cascades in oriented crystals of
  garnet and tungstate},
  \href{https://doi.org/10.1016/S0370-2693(99)00444-X}{\emph{Physics Letters B}
  {\bfseries 456} (1999) 86}.

\bibitem{baryshevsky2017}
V.G.~Baryshevsky et~al., \emph{On the influence of crystal structure on the
  electromagnetic shower development in the lead tungstate crystals},
  \href{https://doi.org/10.1016/j.nimb.2017.02.066}{\emph{Nucl. Instrum.
  Methods Phys. Res. B} {\bfseries 402} (2017) 35}.

\bibitem{bandiera2019}
L.~Bandiera et~al., \emph{Compact electromagnetic calorimeters based on
  oriented scintillator crystals},
  \href{https://doi.org/https://doi.org/10.1016/j.nima.2018.07.085}{\emph{Nucl.
  Instrum. Methods Phys. Res. A} {\bfseries 936} (2019) 124}.

\bibitem{sytov2023}
A.~Sytov et~al., \emph{{Geant4 simulation model of electromagnetic processes in
  oriented crystals for accelerator physics}},
  \href{https://doi.org/10.1007/s40042-023-00834-6}{\emph{J. Korean Phys. Soc.}
  {\bfseries 83} (2023) 132}.

\bibitem{pmgthesis}
P.~Monti-Guarnieri, \emph{{Beamtest characterization of oriented crystals for
  the KLEVER Small Angle Calorimeter}},  2023.

\bibitem{agostinelli2003}
S.~Agostinelli et~al., \emph{{Geant4 -- A simulation toolkit}},
  \href{https://doi.org/10.1016/S0168-9002(03)01368-8}{\emph{Nucl. Instrum.
  Methods Phys. Res. A} {\bfseries 506} (2003) 250}.

\bibitem{pdg}
{\scshape Particle Data Group} collaboration, \emph{{Review of Particle
  Physics}}, \href{https://doi.org/10.1093/ptep/ptac097}{\emph{PTEP} {\bfseries
  2022} (2022) 083C01}.

\bibitem{allison2016}
J.~Allison et~al., \emph{{Recent developments in Geant4}},
  \href{https://doi.org/10.1016/j.nima.2016.06.125}{\emph{Nucl. Instrum.
  Methods Phys. Res. A} {\bfseries 835} (2016) 186}.

\bibitem{guidi2012}
V.~Guidi et~al., \emph{Radiation generated by single and multiple volume
  reflection of ultrarelativistic electrons and positrons in bent crystals},
  \href{https://doi.org/10.1103/PhysRevA.86.042903}{\emph{Phys. Rev. A}
  {\bfseries 86} (2012) 042903}.

\bibitem{bandiera2015}
L.~Bandiera et~al., \emph{{RADCHARM++: A C++ routine to compute the
  electromagnetic radiation generated by relativistic charged particles in
  crystals and complex structures}},
  \href{https://doi.org/https://doi.org/10.1016/j.nimb.2015.03.031}{\emph{Nucl.
  Instrum. Methods Phys. Res. B} {\bfseries 355} (2015) 44}.

\bibitem{sytov2019}
A.I.~Sytov et~al., \emph{Simulation code for modeling of coherent effects of
  radiation generation in oriented crystals},
  \href{https://doi.org/10.1103/PhysRevAccelBeams.22.064601}{\emph{Phys. Rev.
  Accel. Beams} {\bfseries 22} (2019) 064601}.

\bibitem{pedregosa2011}
F.~Pedregosa et~al., \emph{Scikit-learn: Machine learning in {P}ython},
  \href{https://doi.org/arXiv:1201.0490v4}{\emph{J. Mach. Learn. Res.}
  {\bfseries 12} (2011) 2825}.

\bibitem{graczykowski2022}
L.K.~Graczykowski et~al., \emph{{Using machine learning for particle
  identification in ALICE}},
  \href{https://doi.org/10.1088/1748-0221/17/07/C07016}{\emph{J. Instrum.}
  {\bfseries 17} (2022) }.

\bibitem{albert2008}
J.~Albert et~al., \emph{{Implementation of the Random Forest method for the
  Imaging Atmospheric Cherenkov Telescope MAGIC}},
  \href{https://doi.org/10.1016/j.nima.2007.11.068}{\emph{Nucl. Instrum.
  Methods Phys. Res. A} {\bfseries 588} (2008) 424}.

\bibitem{sklearnrfweb}
``{scikit-learn documentation for the Random Forest classifier}.'' available
  at:
  \url{https://scikit-learn.org/stable/modules/generated/sklearn.ensemble.RandomForestClassifier.html},
  2023.

\bibitem{atwood2009}
W.B.~Atwood et~al., \emph{{The Large Area Telescope on the FERMI gamma-ray
  space telescope mission}},
  \href{https://doi.org/10.1088/0004-637X/697/2/1071}{\emph{{ApJ}} {\bfseries
  697} (2009) 1071}.

\end{thebibliography}\endgroup
\end{document}